\documentstyle[epsf,aps]{revtex}
\begin{document}
\newcommand{\be}{\begin{equation}}
\newcommand{\en}{\end{equation}}
\newcommand{\bea}{\begin{eqnarray}}
\newcommand{\ena}{\end{eqnarray}}

\title{Scaling of the distribution of shortest paths in percolation}

\author{Nikolay V. Dokholyan$^{1*}$~\footnotetext{* {\it Correspondence
to:} N. V. Dokholyan. E-mail: dokh@miranda.bu.edu}, Youngki Lee,$^1$
Sergey V. Buldyrev,$^1$\\ Shlomo Havlin,$^2$ Peter R. King$^{3,4}$, and
H. Eugene Stanley$^1$}

\address{$^1$Center for Polymer Studies and Physics Dept., Boston
University, Boston, MA 02215 USA\\ 
$^2$Minerva Center and Department of Physics, Bar-Ilan University, Ramat
Gan, Israel\\ 
$^3$BP Exploration Operating Company Ltd., Sunbury-on-Thames, Middx.,
TW16 7LN, UK \\
$^4$Department of Engineering, Cambridge University, Cambridge, UK.}

\date{{\it J. Stat. Phys.}, {\bf 93:} 603--613 (1998).}

\maketitle

\begin{abstract}

We present a scaling hypothesis for the distribution function of the
shortest paths connecting any two points on a percolating cluster which
accounts for {\it (i)} the effect of the finite size of the system, and
{\it (ii)} the dependence of this distribution on the site occupancy
probability $p$. We test the hypothesis for the case of two-dimensional
percolation.

\end{abstract}


{Dedicated to Professor Leo Kadanoff on the occasion of his
$60^{\mbox{th}}$ birthday.} 

\bigskip


The {\it chemical distance or minimal path}, $\ell$, between two sites
is defined as the shortest path on a percolating cluster connecting the
two sites.  The quantity of interest here is the conditional
probability, $P(\ell|r)$, that two sites taken from the same cluster,
separated by geometrical distance $r$, are $\ell$ chemical distance
away.  For example, in oil recovery the first passage time from the
injection well to a production well a distance $r$ away is related to
$P(\ell|r)$. In many realistic problems where the disordered media
controls a transport process, dynamic properties such as conductivity
and diffusion can also be expressed in terms of chemical distance,
in the case of loopless aggregates or aggregates for which loops can be
neglected \cite{havlin87}.  It is known that the average chemical
distance $\bar\ell$ scales as $r^{d_{\mbox{\scriptsize min}}}$, where
various estimates of $d_{\mbox{\scriptsize min}}$ are
$d_{\mbox{\scriptsize min}}\approx 1.130 \pm 0.005$ \cite{herrmann88}
and $d_{\mbox{\scriptsize min}}\approx 1.1307 \pm 0.0004$
\cite{grassberger92}. There has been an extensive theoretical and computer
work done on studying the scaling of $P(\ell|r)$
\cite{neumann88,Bunde92,Hovi97}, but the complete scaling form of
$P(\ell|r)$ which accounts for the finite size effect, and off-critical
behavior has not been studied in detail.


We start by recalling the scaling form for a different but related
function, the conditional probability $P(r|\ell)$, the probability that
two sites, separated by chemical distance $\ell$, are a geometric
distance $r$ away.  For isotropic media this probability distribution
was studied extensively (see
\cite{havlin87,neumann88,Bunde92,Hovi97,ray85,barma86}). 
In analogy with the theory of self-avoiding random walks (SAWs)
\cite{degennes79,domb69,fisher66}, it was proposed \cite{havlin87} that
\be
P(r|\ell)=A_l \left({r \over \ell^{\tilde{\nu}}}\right)^{g_r}
f_0\left(\frac{r}{\ell^{\tilde{\nu}}}\right), 
\label{eq:prl}
\en
where $A_\ell \sim 1/\ell^{\tilde{\nu}}$,
$\tilde{\nu}=1/d_{\mbox{\scriptsize min}}=0.88\pm 0.02$, $g_r=2.2 \pm
0.3$ \cite{porto97} and the scaling function
$f_0(x)=\exp(-ax^{\tilde{\delta}})$ with $\tilde{\delta} =
(1-\tilde{\nu})^{-1}$.  The relation between exponent $g_r$ and other
known exponents was proposed by Havlin \cite{havlinU}, who used
arguments similar to those of de~Gennes \cite{degennes79} for SAWs. The
distribution $P(r|\ell)_{r=1}$ determines the probability of the minimal
path returns to a nearest neighbour site of the origin, which is from
Eq.~(\ref{eq:prl})
\be
{1\over\ell^{\tilde\nu}}\left({1\over\ell^{\tilde\nu}}\right)^{g_r}=
{1\over\ell^{\tilde\nu+\tilde\nu g_r}}.
\label{eq:prla}
\en
On the other hand, the total number of clusters $N(\ell)$ of chemical
size $\ell$ scales as $M^{-\tau+1}(dM/d\ell)$, where $M$ is the total
mass of the cluster, which scales as $\ell^{\tilde\nu d_f}$, where $d_f$
is the fractal dimension of the percolation cluster, and
$\tau=1+d/d_f$ (see e.g. \cite{havlin87}). Hence
\be
N(\ell)\sim\ell^{\tilde\nu d_f(-\tau+1)+\tilde\nu d_f-1}\sim
\ell^{\tilde\nu d_f(-\tau+2)-1}.
\label{eq:prlb}
\en
Due to the fact that in the mean field regime the effect of the excluded
volume effect is absent, we conjecture\footnote{This conjecture is made
in analogy with the number of self-avoiding random walks coming to the
origin, which scaling lacks prefactor $N^{\gamma-1}$, i. e. $\gamma=1$ as
in mean field regime (see \cite{degennes79}).} that the number of clusters
$N_0(\ell)$ in which the minimum path $\ell$ returns to a nearest
neighbour site of the origin obeys the mean field values for which
$\tilde\nu=1/2$, $\tau=5/2$, and $d_f=4$. The probability of a cluster
with the path returning to a nearest neighbour site of the origin is
proportional to the fraction of paths ${N_0(\ell)/ N(\ell)}$ which
return to a nearest neighbour site of the origin and the probability of
an end of the path in a volume $V$ to be at the specific site
${1/V}$. Thus,
\be
P(r|\ell)_{r=1}\sim {N_0(\ell)\over N(\ell)}{1\over V},
\label{eq:prlc}
\en
where $V$ is the volume of the cluster $V\sim\ell^{d\tilde\nu}$. The
expressions (\ref{eq:prla}) and (\ref{eq:prlc}) should be equal and thus
\be
{1\over\ell^{\tilde\nu+\tilde\nu g_r}}\sim {1\over\ell^{\tilde\nu
d_f(-\tau+2)+1+d\tilde\nu}}.
\label{eq:prld}
\en
Finally we have
\be
g_r=d_f(-\tau+2)+d_{\mbox{\scriptsize min}}+d-1=d_f+d_{\mbox{\scriptsize
min}}-1. 
\label{eq:prle}
\en
This value is within the error bars found numerically for $g_r$ in $d=2$
and $d=3$ \cite{schwarzer98}.  Numerical simulations for percolation
clusters in $d=2$ are in good agreement with the above scaling form
\cite{havlin87,neumann88,bunde96}.

We start by recalling a scaling Ansatz for $P(\ell|r)$, which has been
developed in \cite{neumann88,Bunde92,Hovi97}. Exactly at $p=p_c$ in the
infinite system ($L=\infty$), in analogy to (\ref{eq:prl}),
\be
P(\ell|r) = A_r \left({\ell \over 
r^{d_{\mbox{\scriptsize min}}}}\right)^{-g_{\ell}} f_1\left({\ell \over 
r^{d_{\mbox{\scriptsize min}}}}\right)\,         \label{eq:plr}
\en
where $f_1(x) \sim \exp(-ax^{-\phi_1})$ is the scaling function
corresponding to $f_0$ and $A_r\sim {1\over r^{d_{\mbox{\scriptsize
min}}}}$ is the normalization factor. The conditional probabilities
$P(r|\ell)$ and $P(\ell|r)$ are related by
\be
P(r|\ell)P(\ell)=P(\ell|r)P(r)=P(r,\ell)\, , \label{eq:cond_prob}
\en
where $P(r)=\int P(r,\ell)d\ell$ and $P(\ell)=\int P(r,\ell) dr$.  For
the case of the infinite percolation cluster, i. e. when we restrict
ourself only to the points on the infinite cluster, $P(r)$ and $P(\ell)$
scale as \cite{bunde96}
\be 
P(\ell)\sim \ell^{d_{\ell}-1}\, ,\, P(r)\sim r^{d_f-1}\, ,
\en
where $d_{\ell}=d_f/d_{\mbox{\scriptsize min}}$. Thus,
\be
P(\ell|r) \sim P(r|\ell) {\ell^{d_{\ell}-1} \over r^{d_f-1}}\,
.\label{eq:rel}
\en
Substituting Eqs.~(\ref{eq:prl}) and (\ref{eq:plr}) into
Eq.~(\ref{eq:rel}), we find that
\be
f_1(x)\sim \exp(-ax^{-\phi_1})\, 
\en
where
\be
\phi_1=\tilde\delta\tilde\nu={\tilde\nu (1-\tilde\nu)^{-1}}= 
1/(d_{\mbox{\scriptsize min}}-1),
\label{eq:phi1}
\en
and \cite{bunde96}
\be
g_{\ell}-1=(g_r-1)\tilde\nu+(2-d_f)\tilde{\nu}.
\label{eq:8}
\en
Using Eq.~(\ref{eq:prle}) we find $g_{\ell}=2$ for all $d\ge 2$. Note,
that the numerical value for $g_{\ell}\approx 2.04$ found here is very
close to this prediction.

Let us now consider the scaling of $P'(\ell|r)$, which is defined in
the same way as $P(\ell|r)$, but for any two randomly chosen points,
separated by geometrical distance $r$ and not necessarily belonging to
an infinite cluster. In this case,
\be
P'(r)\sim r G(r)\, , \label{eq:P_G}
\en
where $G(r)$ is a pair connectedness function and the factor $r$ comes
from the summation over all points which are equidistant from the
origin.  The scaling of the $G(r)$ is described in \cite{stanley71}. At
$p=p_c$:
\be
G(r)\sim r^{2-d-\eta}\, ,\label{eq:G_r}
\en
where $\eta =4-2d_f$. The exact value of $\eta = 5/24\approx 0.21$ is
known for $d=2$. Substitution of Eqs.~(\ref{eq:plr}) and (\ref{eq:P_G})
into Eq.~(\ref{eq:cond_prob}), and normalization, yields
\be
P'(r,\ell) \sim r G(r) {1\over {r^{d_{\mbox{\scriptsize min}}}}}
\left(\frac{\ell}{r^{d_{\mbox{\scriptsize min}}}}\right)^{-g'_l}
f_1\left(\frac{\ell}{r^{d_{\mbox{\scriptsize min}}}}\right)
\sim  r^{1-d_{\mbox{\scriptsize min}}-\eta}
\left(\frac{\ell}{r^{d_{\mbox{\scriptsize min}}}}\right)^{-g'_l}
f_1\left(\frac{\ell}{r^{d_{\mbox{\scriptsize min}}}}\right).
\label{eq:Plr1}
\en
The integration of $P'(r,\ell)$ over $r$ gives
\be
P'(\ell)\sim \ell^{2\tilde{\nu}-1-\tilde{\nu}\eta}\, .\label{eq:pl}
\en
Next, we substitute Eqs.~(\ref{eq:Plr1}) and (\ref{eq:pl}) to
Eq.~(\ref{eq:cond_prob}) and compare with (\ref{eq:prl}). We obtain
\be
P'(\ell|r) = A_r \left({\ell \over r^{d_{\mbox{\scriptsize
min}}}}\right)^{-g_{\ell}'} f_1\left({\ell \over r^{d_{\mbox{\scriptsize
min}}}}\right),
\en
where
\be
g_{\ell}'-1=(g_r-1)\tilde\nu+2(2-d_f)\tilde{\nu}.
\label{eq:14}
\en
Thus from (\ref{eq:8}) and (\ref{eq:14}),
\be
g_{\ell}'-g_\ell=(2-d_f)\tilde\nu.
\en
The latter relation between exponents $g_{\ell}'$ and $g_{\ell}$ has
simple probabilistic meaning, since a pair of two randomly selected
points separated by a chemical distance $\ell$ should belong to the
cluster of chemical size $\ell_0 > \ell$. The probability of this event
scales as $\ell^{-\eta\tilde{\nu}/2}$. Once two points belong to such a
cluster, the probability that their chemical distance is equal to $\ell$
scales the same way as on an infinite cluster  and is proportional to
$\ell^{-g_\ell}$. Hence the probability that two randomly selected
points are separated by a chemical distance $\ell$ is proportional to
the product of these two probabilities $\ell^{-g_\ell-\eta\tilde\nu/2}$,
which, by definition, is $\ell^{-g_\ell'}$. Hence
$g_\ell'=g_\ell+\eta\tilde\nu/2$.\footnote{Indeed, $\eta\tilde\nu/2 =
\beta\tilde\nu/\nu$, where $\beta$ is defined as the exponent of the
probability to belong to the infinite cluster, $P_{\infty}\sim
|p-p_c|^{\beta}$. Analogous relation is known, for example, for the
exponents $\gamma$ and $1/\sigma$, which define respectively the scaling
of the average and the typical size of the cluster:
$1/\sigma=\gamma+\beta$.} We summarize the values of the above exponents
for $d=2$ in Table I.

Next we study the finite size effect and off-critical behavior
using Eq. (\ref{eq:plr}) as starting point. We propose that for $\xi >
r$, $P'(\ell|r)$ has the following scaling form
\be
P'(\ell|r) \sim {1 \over r^{d_{\mbox{\scriptsize min}}}}\left({\ell \over
r^{d_{\mbox{\scriptsize min}}}}\right)^{ -g_{\ell}'}
          f_1\left({\ell \over r^{d_{\mbox{\scriptsize min}}}}\right)~
          f_2\left({\ell \over L^{d_{\mbox{\scriptsize min}}}}\right)~
          f_3\left({\ell \over \xi^{d_{\mbox{\scriptsize min}}}}\right)\, , 
\label{eq:ansatz}
\en
where the scaling functions are $f_1(x)=\exp(-ax^{-\phi_1})$, $f_2(x)
=\exp(-bx^{\phi_2})$ and $f_3(x)=\exp(-cx)$.  Here $\xi$ is a
characteristic length of pair connectedness function and has a power-law
dependence of occupancy $p$ as
\be
\xi \sim |p-p_c|^{-\nu}.  \label{eq:corr}
\en
The first function $f_1$ accounts for the lower cut-off due to the
constraint $\ell>r$, while $f_2$ and $f_3$ account for the upper cut-off
due to the finite size effect and due to the finite correlation length
respectively.  Either $f_2$ and $f_3$ becomes irrelevant, depending on
which of the two values $L$ or $\xi$ is greater. For $L<\xi$, $f_2$
dominates the upper cut-off, otherwise $f_3$ dominates. We assume the
independence of the finite size effect and the effect of the
concentration of the vacant sites, so the Eq.~(\ref{eq:ansatz})
can be represented as a product of the terms which are responsible for
the finite size effect ($f_2$) and the effect of the concentration
($f_3$). Our simulations confirm this assumption numerically.

There is no particular reason for the choice of the simple exponential
decay of the function $f_3$, but we find that this function is a
straight line on a semilogarithmic plot (see discussion below). However,
due to the fact that the  pair connectedness function $G(r)\sim
r^{2-d-\eta} \exp{(-r/\xi)}$ has simple exponential form
\cite{stanley71} it is reasonable to assume that function $f_3$ has a
form of a stretched exponential $f_3\sim
\exp(-cx^{\tilde\nu})$. However, since the exponent $\tilde\nu\approx
0.88$ is close to 1, we are not able to resolve this question in our
simulations.



Next, using the Leath method \cite{leath76,alexandrowicz81,pike81},
which corresponds to calculation of $P'(\ell,r)$ \footnote{
In the Leath method, starting a cluster from the origin
corresponds to selection of a random point in the disordered
media. Construction of the Leath cluster corresponds to the
determination of all the points which are connected to the starting
one. In our procedure, we compute $P'(\ell,r)$ by dividing the number of
the pair of points $N_{\ell,r}$ which are separated by $r$ and $\ell$ by
the total number of points $N$ in all constructed clusters. The quantity
of interest  $N_{\ell,r}/N$ is equivalent to the quantity $P'(\ell,r)$,
which by definition is the probability that two randomly chosen points,
belonging to the same cluster, are separated by $r$ and $\ell$.}, we
numerically test the scaling conjecture Eq.~(\ref{eq:ansatz}) exactly at
the percolation threshold $p=p_c$. In this case, $\xi=\infty$ so
$f_3=\mbox{const}$. Hence Eq.~(\ref{eq:ansatz}) reduces to
\be
P'(\ell|r) \sim {1 \over r^{d_{\mbox{\scriptsize min}}}} \left({\ell \over
r^{d_{\mbox{\scriptsize min}}}}\right)^{-g_{\ell}'} f_1\left({\ell\over
r^{d_{\mbox{\scriptsize min}}}}\right)~ f_2\left({\ell\over
L^{d_{\mbox{\scriptsize min}}}}\right), ~~~ (p=p_c).
\label{eq:plr1}
\en
Indeed, Fig.~1a shows that $P'(\ell|r)$ has power law behavior for
$r^{d_{\mbox{\scriptsize min}}}<\ell<L^{d_{\mbox{\scriptsize min}}}$ and
rapidly vanishes for $\ell<r^{d_{\mbox{\scriptsize min}}}$ and for
$\ell>L^{d_{\mbox{\scriptsize min}}}$. In order to test Eq.
({\ref{eq:plr1}) we compute rescaled probability distribution
\be
\Phi\left({\ell\over r^{d_{\mbox{\scriptsize min}}}}\right) 
\equiv P'(\ell|r)\ell^{g_{\ell}'}r^{-d_{\mbox{\scriptsize min}}(g_{\ell}'-1)}
=f_1 \cdot f_2,
\label{eq:rescaled} 
\en
and plot it against scaling variable $x=\ell/r^{d_{\mbox{\scriptsize
min}}}$ (see Fig.~1b). According to Eq.~(\ref{eq:plr1}),
\be
\Phi(x)=Af_1(x)f_2\left[ x\left({r\over L}\right)^{d_{\mbox{\scriptsize
min}}}\right]. \label{eq:Phix}
\en
Therefore, $\Phi(x)$ should depend only on $x$ and the ratio $L/r$.
Indeed, Fig.~1b shows excellent data collapse for $L/r=16$, with sharp
cutoffs governed for small $x<1$ by $f_1(x)$ and for large
$x>(L/r)^{d_{\mbox{\scriptsize min}}}$ by
$f_2[x(r/L)^{d_{\mbox{\scriptsize min}}}]$.

In order to test the assumption that the functions $f_1$ and $f_2$ are
stretched exponentials with exponents $\phi_1$ and $\phi_2$ we plot
$\Pi(x)=\log[A/\Phi(x)]$ versus $x$ in double logarithmic scale for
various values of normalization constant $A$. (See Fig.~2.)  If the
stretched exponential conjecture is correct, $\Pi(x)$ should have two
straight line asymptotes for $\log x \rightarrow \infty$ with the slope
$\phi_2$ and for $\log x \rightarrow -\infty$ with the slope $-\phi_1$.
The slopes $\phi_1$ and $\phi_2$ of the straight line fits depend weakly
on the value of $A$.  Using $A=1.65$, we obtain the longest regimes of
straight line behavior. For this $A$ we obtained $\phi_1 \approx 7.3$
and $\phi_2\approx 4.0$.  The value of $\phi_1$ is in good agreement
with Eq.(\ref{eq:phi1}), while $\phi_2$ seems to be a new exponent.

Finally, in order to test the dependence of $P'(\ell|r)$ on $p$ we obtain
data for very large system size $L$ and for several values of $p\neq
p_c$. In this case, the upper cutoff of the distribution
Eq.~(\ref{eq:ansatz}) is governed by $f_3$ and the functional form of
the rescaled probability $\Phi$ is given by
\be
\Phi(\ell/r^{d_{\mbox{\scriptsize min}}})
\sim f_1\left({\ell\over r^{d_{\mbox{\scriptsize min}}}}\right)
f_3\left({\ell\over\xi^{d_{\mbox{\scriptsize min}}}}\right). \label{eq:phix}
\en    
For large $\ell$, we suggest an exponential decay of $\Phi$:
\be
\Phi(\ell/r^{d_{\mbox{\scriptsize min}}})
\sim \exp\left(-c{\ell\over\xi^{d_{\mbox{\scriptsize
min}}}}\right). 
\label{eq:plrp}
\en
Indeed, semi-logarithmic plots of $\log
\Phi(\ell/r^{d_{\mbox{\scriptsize min}}})$ versus $\ell$
shown in Fig.~3a can be approximated by the straight lines with slopes
which approach to zero as $p \rightarrow p_c$.
According to Eq.~(\ref{eq:plrp}), these slopes $k(p)$
should be proportional to $\xi^{-d_{\mbox{\scriptsize
min}}}=|p-p_c|^{d_{\mbox{\scriptsize min}}\nu}\approx |p-p_c|^{1.51}$.
Fig.~3b shows double logarithmic plots of $|k(p)|$ versus $|p-p_c|$
for $p<p_c$ and for $p>p_c$, which can be well approximated by straight
lines with slopes $1.55$ and $1.57$ in good agreement with scaling
conjecture, Eq.~(\ref{eq:ansatz}).

The scaling form, Eq.(\ref{eq:ansatz}), is limited to the case when
$\xi>r$.  For $\xi<r$, the finite size effects can be neglected, the
power law regime vanishes, and the minimal path can be divided into
$r/\xi$ independent blobs each of length $\xi^{d_{\mbox{\scriptsize
min}}}$, so that the distribution $P'(\ell|r)$ approaches Gaussian form
with mean $r\xi^{d_{\mbox{\scriptsize min}}-1}$ and variance
$r\xi^{2d_{\mbox{\scriptsize min}}-1}$.


In many realistic problems an interesting quantity is the first passage
time distribution.  If we assume that the first passage time $t$ between
two points is linearly proportional to shortest chemical distance $\ell$
between them then we can identify the distribution $P'(\ell|r)$ as a
first passage time distribution.  This is only true when the spreading
velocity through the media remains constant along the path.  In the oil
recovery problem it is the injection rate, the volume of the fluid
pumped into the media per unit time, which is maintained constant, not
the velocity itself.  We can generalize the relation between first
passage time and chemical distance by $t \sim \ell^z$.  For the given
fractal dimension $d_B$ of the backbone of the percolating cluster we
know that $M(r) \sim r^{d_B}$ where $M(r)$ is the total mass of the
backbone of the percolation cluster of radius $r$.  Constant pumping
rate gives $dt\sim dM\sim r^{d_B-1}dr$.  Combining with the relation
$\ell \sim r^{d_{\mbox{\scriptsize min}}}$ and their derivative $d\ell
\sim r^{d_{\mbox{\scriptsize min}}-1} dr$ we can get another scaling
relation $z=d_B/d_{\mbox{\scriptsize min}}=1.625/1.14\approx 1.43$. If
we substitute $\ell \sim t^{1/z}$ in Eq.~(\ref{eq:ansatz}) it gives the
correct scaling form for the first passage time distribution in case of
constant injection rate.

Another problem comes from the presence of a second well,
which might establish some pressure potential field through the media
and bring some anisotropic spreading front toward the production well
when we apply some external driving pressure.  This can be corrected by
calculating the potential field which is a solution of Laplace's
equation with two point boundaries and calculating the local velocity on
the percolating cluster including the quenched disorder of the
percolation cluster.  Practically the problem of multiple wells
scattered along the oil fields with equal spacing is of particular
interest in oil recovery.

In summary, we have studied the scaling properties of the shortest paths
distribution for fixed two points on the percolating cluster which
accounts for the finite size effect, also off criticality.  We propose a
plausible scaling hypothesis for the distribution, which is supported by
theoretical argument and tested by numerical simulation.  The lower and
upper cut-offs of the distribution has been numerically observed and
fitted successfully by stretched exponential function.  Off the critical
point the upper cutoff, due to the finite correlation length, becomes a
pure exponential form. Further work is underway to test the conjectures
and hypothesis developed here.

We thank Prof. D. Stauffer for fruitful discussion, the referee for the
helpful suggestions, and BP for the financial support.

\begin{center}
\begin{tabular}{|c|c|c|c|c|c|c|c|c|c|}\hline \label{tab:exponents}
$\beta$   &$\nu$       &$\eta$    &$d_{\mbox{\scriptsize min}}$  &$\tilde{\nu}$
&$g_r$    &$g_{\ell}$  &$g_\ell'$ &$\phi_1$  &$\phi_2$ \\\hline
$0.14$    &$1.33$      &$0.20$    &$1.13$    &$0.88$
&$2.2$    &$2.04$      &$2.14$    &$7.3$     &$4.0$  \\\hline
\end{tabular}
\end{center}
Table I: Numerical estimates for the various exponents.

\begin{figure}[htb]
\centerline{
\epsfxsize=8.0cm
\epsfbox{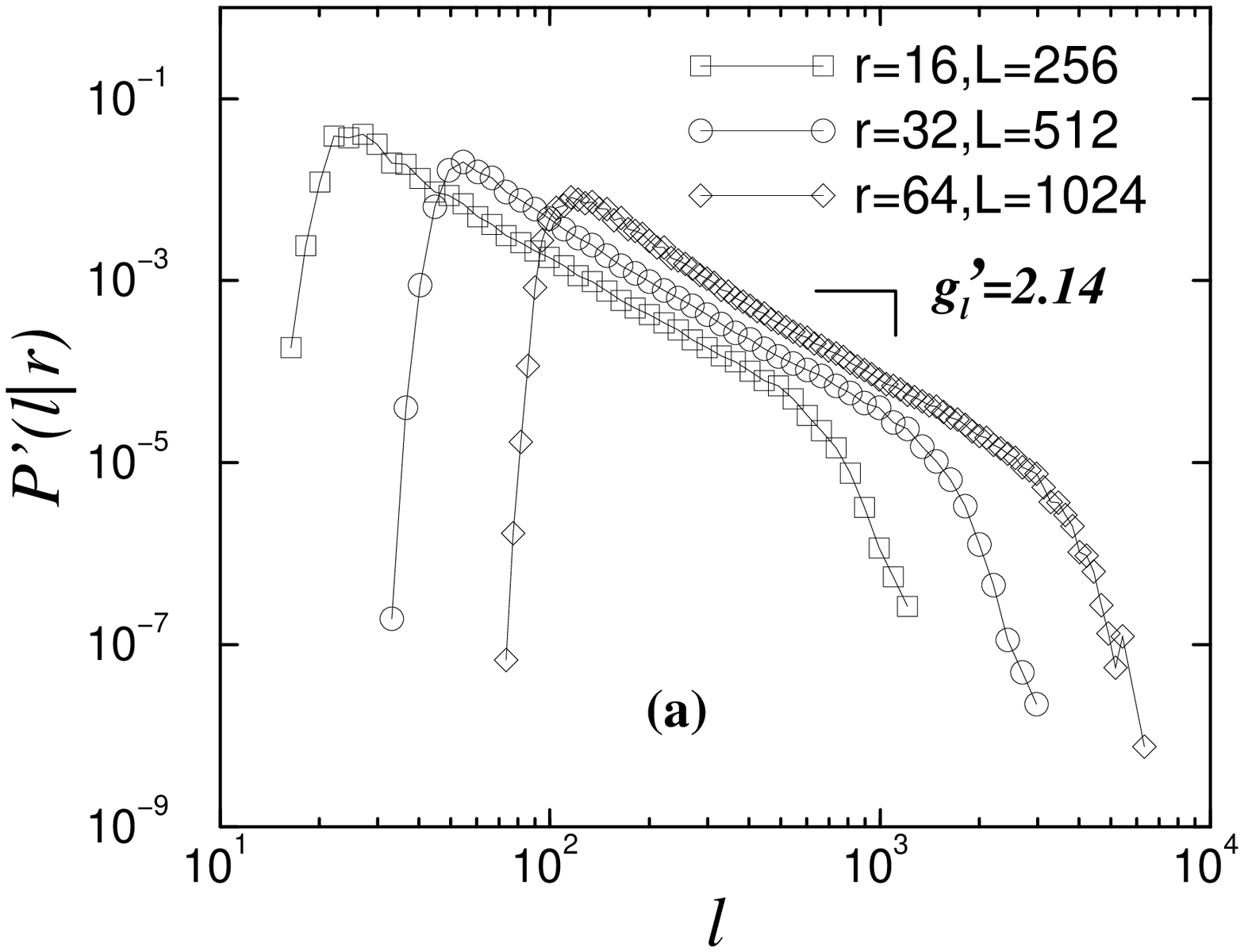} 
\vspace*{1.0cm}
\epsfxsize=8.0cm
\epsfbox{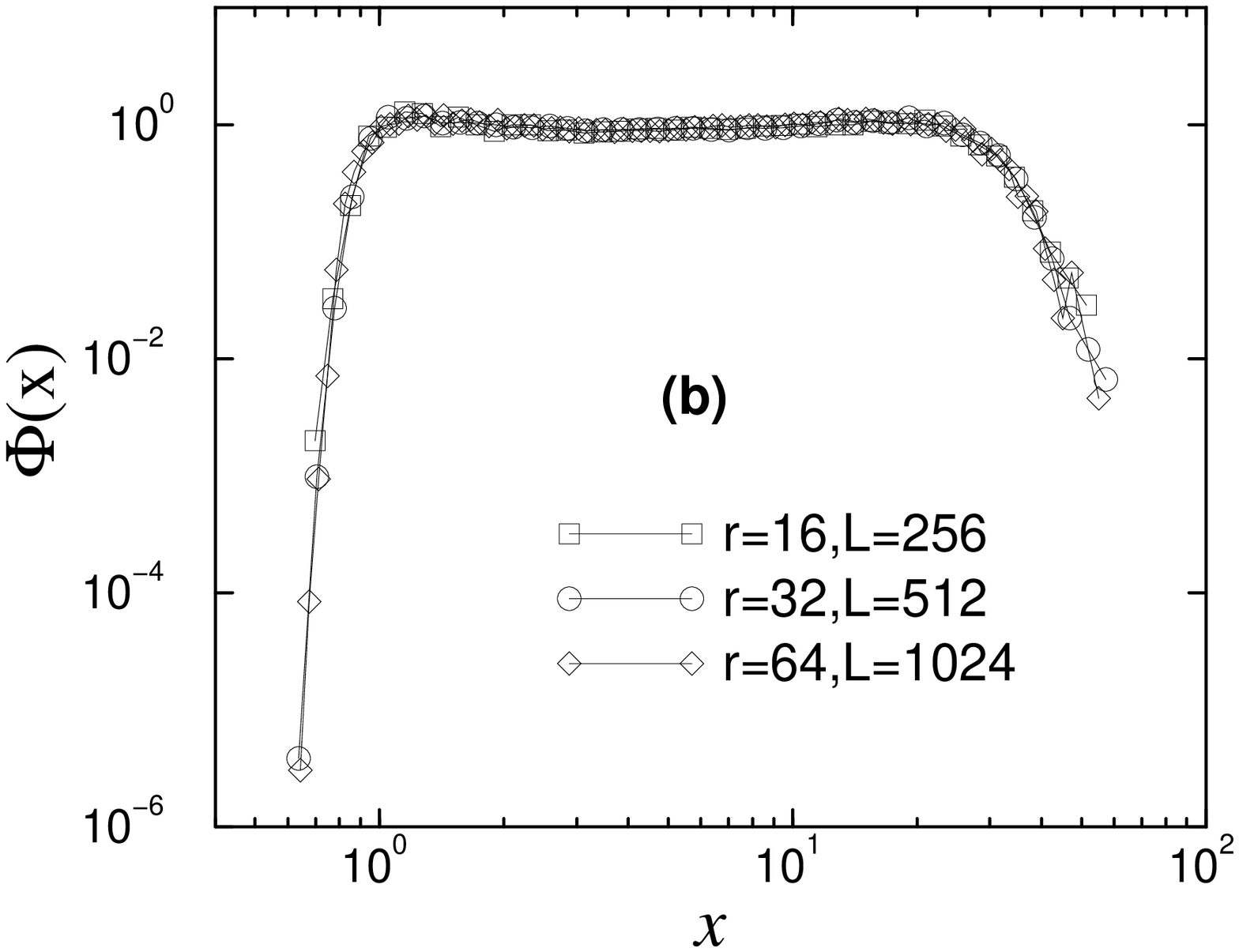} 
\vspace*{1.0cm}
}
\caption{
(a) Log-log plot of $P'(\ell|r)$ for $p=p_c=0.593$ and for different sets
of parameters, $(r,L)$.  The straight line regime has slope
$g_{\ell}'=2.14$.  (b) Log-log plot of rescaled probability
$\Phi(x)\equiv P'(\ell |r)\ell^{g_{\ell}'}r^{-d_{\mbox{\scriptsize
min}}(g_{\ell}'-1)}$ against rescaled length
$x=\ell/r^{d_{\mbox{\scriptsize min}}}$ using the values,
$g_{\ell}'=2.14$ and $d_{\mbox{\scriptsize min}}=1.14$.}
\label{fig:1}
\end{figure}

\begin{figure}[htb]
\centerline{
             \epsfxsize=8.0cm
\epsfbox{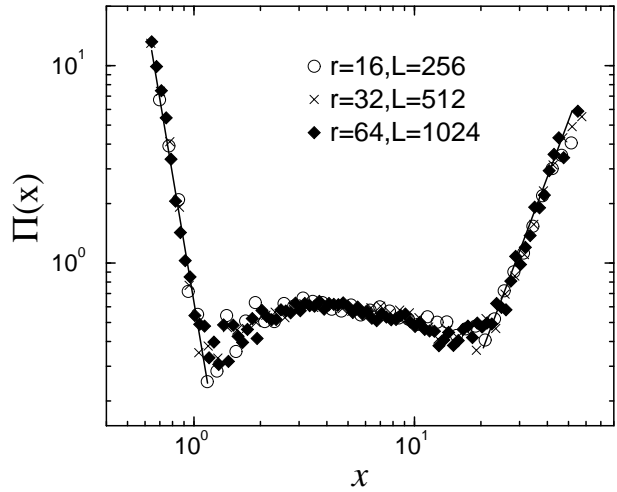} 
\vspace*{1.0cm}
}
\caption{
Log-log plot of transformed probability $\Pi(x)=log[A/\Phi(x)]$ versus
$x=\ell/r^{d_{\mbox{\scriptsize min}}}$.  The slopes of the solid lines
give the power of the stretched exponential function $f_1$ and $f_2$ in
Eq. (\protect{\ref{eq:Phix}}).  Using the parameter $A=1.65$, the slopes
give $\phi_1 \approx 7.3$ for the lower cut-off and $\phi_2 \approx 4.0$
for the upper cut-off.
}
\label{fig:2}
\end{figure}

\begin{figure}[htb]
\centerline{
\epsfxsize=8.0cm
\epsfbox{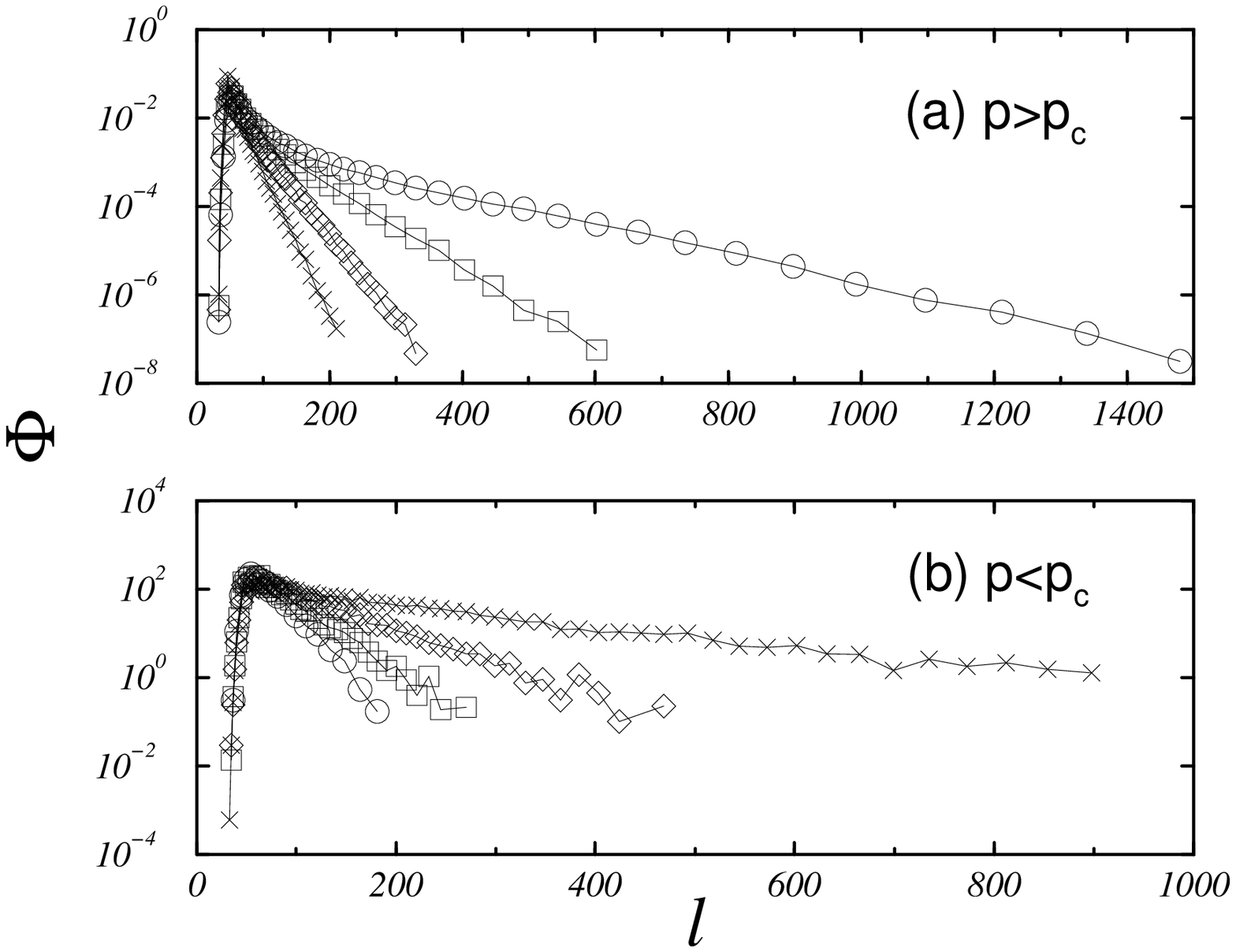} 
\vspace*{1.0cm}
\epsfxsize=8.0cm
\epsfbox{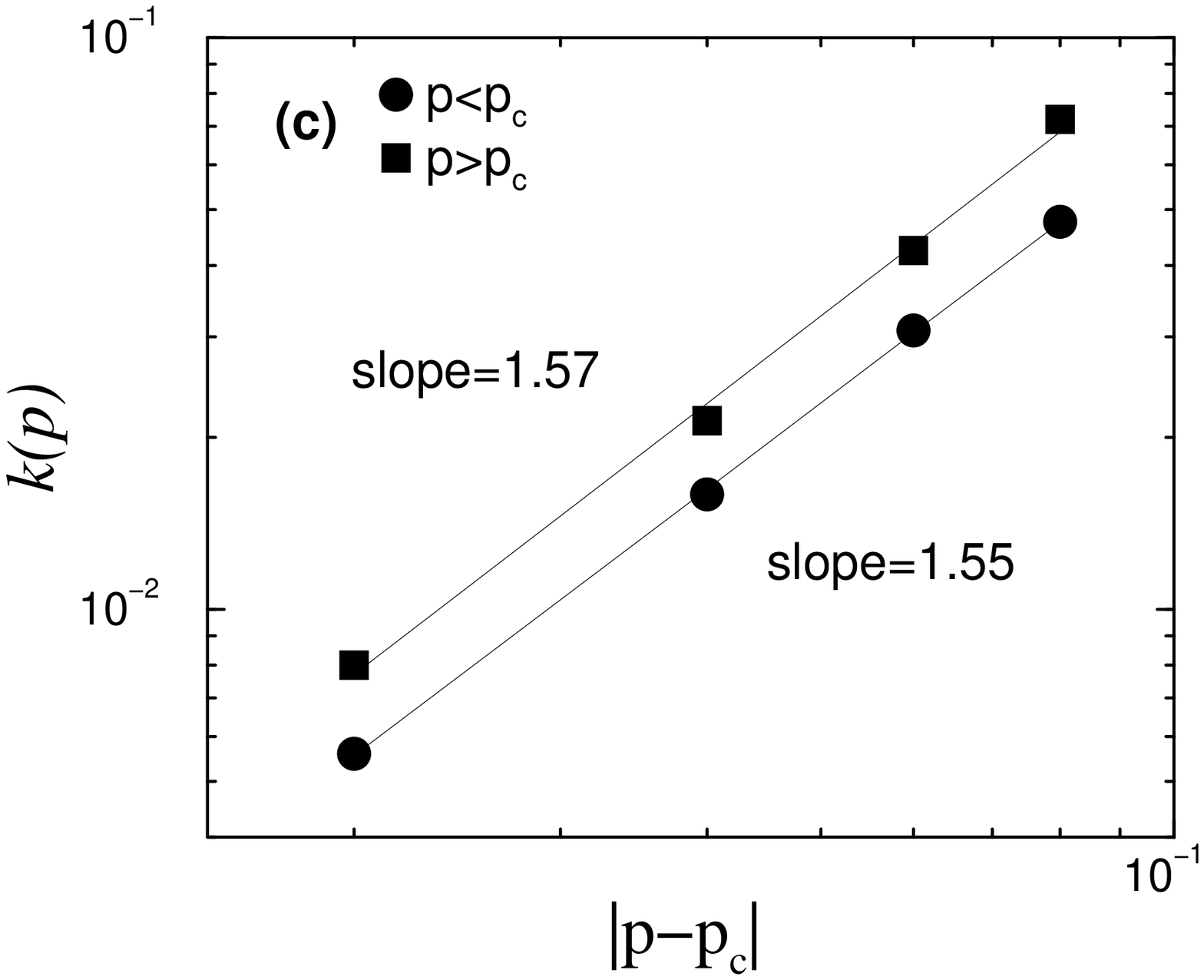} 
\vspace*{1.0cm}
}
\caption{
Semi-logarithmic plot of transformed probability
$\Phi$ versus $\ell$ ($r=32$) shows pure exponential behavior of $f_3$.
(a) ($\bigcirc$) denotes $p=0.613$, ($\Box$) $p=0.633$, ($\Diamond$)
$p=0.655$ and ($\times$) $p=0.673$.  (b) ($\bigcirc$) denotes $p=0.533$,
($\Box$) $p=0.553$, ($\Diamond$) $p=0.573$ and ($\times$) $p=0.583$.
(c) The slope of the log-log plot of the coefficient in
exponential function $f_3$ as a function of $|p-p_c|$ gives the value
$\nu d_{\mbox{\scriptsize min}} \approx 1.55$ for $p>p_c$ and $1.57$ for
$p<p_c$.}
\label{fig:3}
\end{figure}

\end{document}